\documentstyle[12pt,epsfig]{article}
\begin{document}
\begin{titlepage}
\title{Effective Field Theory and $\chi$pt}
\author{Barry R. Holstein\\[5mm]
Department of Physics and Astronomy\\University of Massachusetts\\
Amherst, MA  01003}
\maketitle
\begin{abstract}
A brief introduction to the subject of chiral perturbation theory 
($\chi$pt) is given, including
a discussion of effective field theory and application to the 
upcoming Bates virtual Compton scattering measurement.
\end{abstract}
\end{titlepage}
\section{Introduction}
We have gathered 
to celebrate the fact that Bates has been delivering beam successfully for 
twenty five years and to review some of the things which have been learned 
and which are still to be studied.  One thing that {\it has} changed 
theoretically during this period is that we now have a new paradigm for
analysis of low energy processes such as studied at Bates.  I was a 
student in the 1960's and at that time our goal was to attempt to
find a renormalizable field theory which describes all particle interactions 
with the same sort of success as quantum electrodynamics (QED).  In 1967
we went part of the way with development of the Weinberg-Salam theory, which
incorporated the weak interaction as a sibling to the electromagnetic.
Because the interaction was weak it could be treated via the same perturbative
techniques as could its electromagnetic kin and what has resulted is an
extremely successful description of all weak and electromagnetic processes.

For the strong interactions a renormalizable picture has
also been developed---quantum chromodynamics or QCD. The theory is, of course,
deceptively simple on the surface.
Indeed the form of the Lagrangian\footnote{Here the covariant derivative is
\begin{equation}
i D_{\mu}=i\partial_{\mu}-gA_\mu^a {\lambda^a \over 2} \, ,
\end{equation}
where $\lambda^a$ (with $a=1,\ldots,8$) are the SU(3) Gell-Mann matrices,
operating in color space, and the color-field tensor is defined by
\begin{equation}
G_{\mu\nu}=\partial_\mu  A_\nu -  \partial_\nu  A_\mu -
g [A_\mu,A_\nu]  \, ,
\end{equation} }
\begin{equation}
{\cal L}_{\mbox{\tiny QCD}}=\bar{q}(i  {\not\!\! D} - m )q-
{1\over 2} {\rm tr} \; G_{\mu\nu}G^{\mu\nu} \, .
\end{equation}
is elegant, and the theory is renormalizable.  So why are we not
satisfied?  While at the very largest energies, asymptotic freedom 
allows the use of perturbative
techniques, for those who are interested in making contact with low energy 
experimental findings there exist at least three fundamental difficulties:
\begin{itemize}
\item [i)] QCD is written in terms of the "wrong" degrees of 
freedom---quarks and
gluons---while low energy experiments are performed with hadronic bound states;

\item [ii)] the theory is non-linear due to gluon self interactions;

\item[iii)] the theory is one of strong coupling---$g^2/4\pi\sim 1$---so that 
perturbative methods are not practical.
\end{itemize}
Nevertheless, there has been a great deal of recent progress in 
making contact between
theory and experiment using the technique of "effective field theory'', 
which exploits the 
chiral symmetry of the QCD interaction.  In order to understand how 
this is accomplished, we shall first review this idea of effective field theory in
the simple context of quantum mechanics.  Then
we show how these ideas can be married via chiral perturbation theory 
and indicate applications at Bates.

\section{Effective Field Theory}

The power of effective field theory is associated with the feature that there
exist many situations in physics involving {\it two scales}, one heavy and one 
light.  Then, provided one is working at energies small compared to
the heavy scale, it is possible to 
fully describe the interactions in terms of an ``effective'' picture,
which is written only in terms of the light degrees of freedom, but which
fully includes the 
influence of the heavy mass scale through virtual effects.
A number of very nice review articles on effective field theory can be
found in ref. \cite{eftr}.
 
Before proceeding to QCD, however, it is useful to study this idea 
in the simpler context of ordinary quantum mechanics, in order to get familiar 
with the concept.
Specifically, we examine the question of why the sky is blue, whose 
answer can be
found in an analysis of the scattering of photons from the sun 
by atoms in the atmosphere---Compton scattering\cite{skb}.  First we examine 
the problem using traditional quantum mechanics and consider 
elastic (Rayleigh) scattering from, for simplicity,
single-electron (hydrogen) atoms.  The appropriate Hamiltonian is then
\begin{equation}
H={(\vec{p}-e\vec{A})^2\over 2m}+e\phi
\end{equation}
and the leading---${\cal O}(e^2)$---amplitude for Compton scattering
is found from calculating the diagrams shown in Figure 1, yielding the
familiar Kramers-Heisenberg form
\begin{eqnarray}
{\rm Amp}&=&-{e^2/m\over \sqrt{2\omega_i2\omega_f}}\left[\hat{\epsilon}_i\cdot
\hat{\epsilon}_f^*+{1\over m}\sum_n\left({\hat{\epsilon}_f^*\cdot
<0|\vec{p}e^{-i\vec{q}_f\cdot\vec{r}}|n>
\hat{\epsilon}_i\cdot <n|\vec{p}e^{i\vec{q}_i\cdot\vec{r}}|0>\over 
\omega_i+E_0-E_n}\right.\right.
\nonumber\\
&+&\left.\left.{\hat{\epsilon}_i\cdot <0|\vec{p}e^{i\vec{q}_i\cdot\vec{r}}|n>
\hat{\epsilon}_f^*\cdot <n|\vec{p}e^{-i\vec{q}_f\cdot\vec{r}}|0>\over E_0-\omega_f-E_n}
\right)\right]
\end{eqnarray}
where $|0>$ represents the hydrogen ground state having binding energy
$E_0$. 
\begin{figure}
\centerline{\epsfig{file=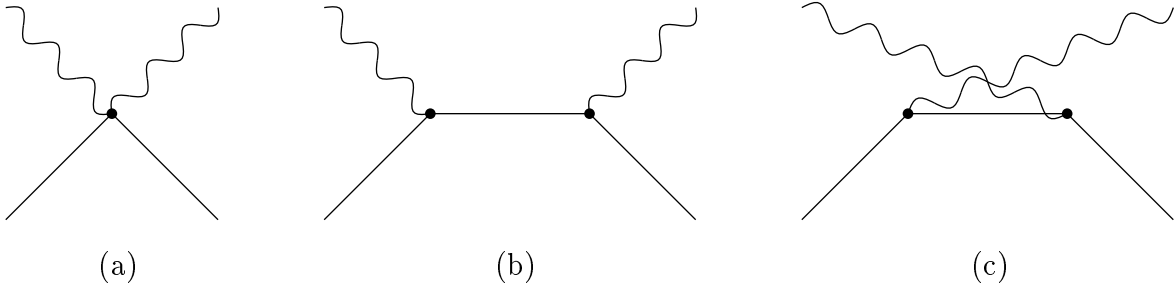}}
\caption{Feynman diagrams for nonrelativistic photonl-atom scattering.}
\end{figure}

Here the leading component is the familiar 
$\omega$-independent Thomson amplitude and would appear naively 
to lead to an energy-independent
cross-section.  However, this is {\it not} the case.  Indeed, by
expanding in $\omega$ and using a few quantum mechanical identities
one can show that, 
provided that the energy of the photon is much smaller than a typical 
excitation energy---as is the case for optical photons, the cross section 
can be written as
\begin{eqnarray}
\quad {d\sigma\over d\Omega}&=&
\lambda^2\omega^4|\hat{\epsilon}_{f}^*\cdot
\hat{\epsilon}_{i}|^2\left(1+{\cal O}\left({\omega^2\over (\Delta E)^2}
\right)\right)\label{eq:cc}
\end{eqnarray}
where 
\begin{equation}
\lambda=\alpha_{em}\sum{2|z_{n0}|^2\over E_n-E_0}
\end{equation}
is the atomic electric polarizability,
$\alpha_{em}=e^2/4\pi$ is the fine structure constant, and $\Delta E\sim m
\alpha_{em}^2$ is a typical hydrogen excitation energy.  We note that  
$\alpha_{em}\lambda\sim a_0^2\times {\alpha_{em}\over \Delta E}\sim a_0^3$ 
is of order the atomic volume, as will be exploited below, and that the cross 
section itself has the characteristic $\omega^4$ dependence which leads 
to the blueness of the sky---blue light scatters much
more strongly than red\cite{feyn}.

Now while the above derivation is certainly correct, it requires somewhat detailed and
lengthy quantum mechanical manipulations which obscure the relatively simple physics
involved.  One can avoid these problems by the use of effective field theory 
methods.  The key point is that of scale.  Since the incident photons
have wavelengths $\lambda\sim 5000$A much larger than the $\sim$ 1A atomic size, then at 
leading order the photon is insensitive to the presence of the atom, since the latter
is electrically neutral.  If $\chi$ represents the wavefunction of the atom then the 
effective leading order Hamiltonian is simply
\begin{equation}
H_{eff}^{(0)}=\chi^*\left({\vec{p}^2\over 2m}+e\phi\right)\chi
\end{equation}
and there is {\it no} interaction with the field.  In higher orders, 
there {\it can} exist such atom-field interactions and this is
where the effective Hamiltonian comes in to play.  In order to construct the
effective interaction, we demand certain general principles---this
Hamiltonian must satisfy fundamental symmetry requirements.  In
particular $H_{eff}$ must be gauge invariant, must be a scalar under
rotations, and must be even under both parity and time reversal 
transformations.  Also,
since we are dealing with Compton scattering, $H_{eff}$ should be
quadratic in the vector potential.
Actually, from the requirement of gauge invariance, it is
clear that the effective interaction can utilize $\vec{A}$ only 
via the electric
and magnetic fields, rather than the vector potential itself---
\begin{equation}
\vec{E}=-\vec{\nabla}\phi-{\partial\over \partial t}\vec{A}, 
\qquad \vec{B}=\vec{\nabla}\times\vec{A}\label{eq:ii}
\end{equation}
since these are invariant under a gauge transformation
\begin{equation}
\phi\rightarrow\phi+{\partial\over \partial t}\Lambda,\qquad \vec{A}
\rightarrow\vec{A}-\vec{\nabla}\Lambda
\end{equation}
while the vector and/or scalar potentials are not.  The lowest order
interaction then can involve only the rotational invariants 
$\vec{E}^2,\vec{B}^2$
and $\vec{E}\cdot\vec{B}$.  However, under spatial
inversion---$\vec{r}\rightarrow -\vec{r}$---electric and magnetic
fields behave oppositely---$\vec{E}\rightarrow -\vec{E}$ while
$\vec{B}\rightarrow\vec{B}$---so that parity invariance rules out any
dependence on $\vec{E}\cdot\vec{B}$.  Likewise under time
reversal invariance $\vec{E}\rightarrow\vec{E},\,\,\vec{B}\rightarrow
-\vec{B}$ so such a term is also T-odd.  
The simplest such effective Hamiltonian must
then have the form
\begin{equation}
H_{eff}^{(1)}=\chi^*\chi[-{1\over 2}c_E\vec{E}^2
-{1\over 2}c_B\vec{B}^2]\label{eq:ll}
\end{equation}
(Terms involving time or spatial derivatives are much smaller.)
We know from electrodynamics that 
${1\over 2}(\vec{E}^2+\vec{B}^2)$
represents the field energy per unit volume, so by dimensional
arguments, in order to represent an
energy in Eq. \ref{eq:ll}, $c_E,c_B$ must have dimensions of volume.
Also, since the photon has such a
long wavelength, there is no penetration of the atom, so  only classical scattering
is allowed.  The relevant scale must then be atomic size so that we can write
\begin{equation}
c_E=k_Ea_0^3,\qquad c_B=k_Ba_0^3
\end{equation}
where we anticipate $k_E,k_B\sim {\cal O}(1)$.  Finally, since for photons
with polarization $\hat{\epsilon}$ and four-momentum $q_\mu$ we
identify $\vec{A}(x)=\hat{\epsilon}\exp(-iq\cdot x)$,
then from Eq. \ref{eq:ii}, $|\vec{E}|\sim \omega$, 
$|\vec{B}|\sim |\vec{k}|=\omega$ and 
\begin{equation}
{d\sigma\over d\Omega}\propto|<f|H_{eff}|i>|^2\sim\omega^4 a_0^6
\end{equation}
as found in the previous section via detailed calculation.  This is a
nice example of the power of simple effective field theory arguments.

\section{Application to QCD: Chiral Perturbation Theory}

Now let's apply these ideas to the case of QCD.  In this case the 
invariance we wish to exploit is  ``chiral symmetry.''   
The idea of "chirality" is defined by the operators
\begin{equation} \Gamma_{L,R} = {1\over 2}(1\pm\gamma_5)={1\over 2}
\left( \begin{array}{c c }
1 & \mp 1 \\
\mp 1 & 1
\end{array}\right)
\end{equation}
which project ``left-'' and ``right-handed'' components of the Dirac 
wavefunction via
\begin{equation} \psi_L = \Gamma_L \psi \qquad \psi_R=\Gamma_R
\psi \quad\mbox{with}\quad \psi=\psi_L+\psi_R \end{equation}
In terms of these chirality states the quark component of the QCD Lagrangian
can be written as
\begin{equation} \bar{q}(i\not\! \! D-m)q=\bar{q}_Li\not \! \! D q_L +
\bar{q}_Ri
\not\!\! D q_R -\bar{q}_L m q_R-\bar{q}_R m
q_L \end{equation}
The reason that these chirality states are called left- and
right-handed is that in the limit $m\rightarrow 0$ they coincide with
quark {\it helicity} projection operators. 
With this background, we note that QCD, in the mathematical limit as  
$m\rightarrow0$ has the structure 
\begin{equation} {\cal L}_{\rm QCD}\stackrel{m=0}{\longrightarrow}
\bar{q}_L i \not\!\! D q_L +
\bar{q}_R i \not\!\! D q_R \end{equation}
and is invariant under {\it independent} global
left- and right-handed rotations
\begin{equation}
q_L  \rightarrow \exp (i \sum_j \lambda_j\alpha_j)
q_L,\qquad
q_R  \rightarrow \exp (i\sum_j \lambda_j \beta_j)
q_R
\end{equation}
This invariance is called
$SU(3)_L \bigotimes SU(3)_R$ or chiral $SU(3)\times SU(3)$.  Continuing
to neglect the light quark masses,
we see that in a chiral symmetric world one would expect to have 
sixteen---eight
left-handed and eight right-handed---conserved Noether currents
\begin{equation} \bar{q}_L\gamma_{\mu} {1\over 2} \lambda_i q_L \, ,
\qquad \bar{q}_R\gamma_{\mu}{1\over 2}\lambda_i
q_R \end{equation}
Equivalently, by taking the sum and difference we would have eight 
conserved vector and
eight conserved axial vector currents
\begin{equation}
V^i_{\mu}=\bar{q}\gamma_{\mu} {1\over 2}
\lambda_i q,\qquad
A^i_{\mu}=\bar{q}\gamma_{\mu}\gamma_5
 {1\over 2} \lambda_i q
\end{equation}
In the vector case, this is just a simple generalization of 
isospin (SU(2)) invariance 
to the case of SU(3).  There exist 
{\it eight} ($3^2-1$) time-independent charges
\begin{equation} F_i=\int d^3 x V^i_0(\vec{x},t) \end{equation}
and there exist various supermultiplets of particles having 
identical spin-parity and
(approximately) the same mass in the configurations---singlet, 
octet, decuplet, 
{\it etc.} demanded by SU(3)-invariance.

If chiral symmetry were realized in the conventional fashion one would
expect there also to exist corresponding nearly degenerate same spin
but {\it opposite} parity states generated
by the action of the time-independent axial charges
$F^{5}_i= \int d^3 xA^i_0(\vec{x},t)$
on these states.  However, it is known that the axial symmetry is
broken spontaneously, whereby Goldstone's theorem requires the
existence of eight massless pseudoscalar bosons, which couple derivatively
to the rest of the universe\cite{goldstone}.  Of course, in the real
world such massless $0^-$ states do not exist, because in the
real world exact chiral invariance is broken by the small quark mass terms
which we have neglected up to this point.  Thus what we have are eight
very light (but not massless) pseudo-Goldstone bosons which make up the
pseudoscalar octet.  Since such states are lighter than their other hadronic
counterparts, we have a situation wherein effective field theory can be 
applied---provided one is working at energy-momenta small compared to 
the $\sim 1$ GeV scale which is typical of hadrons, one can describe the
interactions of the pseudoscalar mesons using an effective Lagrangian.
Actually this has been known since the 1960's, where a good deal of work
was done with a {\it lowest order} effective chiral Lagrangian\cite{gg} 
\begin{equation}
 {\cal L}_2={F_\pi^2 \over 4} \mbox{Tr} (\partial_{\mu}U \partial^{\mu}
 U^{\dagger})+{m^2_{\pi}\over 4} F_\pi^2 \mbox{Tr} (U+U^{\dagger})\,  .\label{eq:abc}
\end{equation}
where the subscript 2 indicates that we are working at two-derivative order
or one power of chiral symmetry breaking---{\it i.e.} $m_\pi^2$.
Here $U\equiv\exp(\sum \lambda_i\phi_i/F_\pi)$, where $F_\pi=92.4$ is the pion
decay constant. This Lagrangian is {\it unique}---if we expand to 
lowest order in $\vec\phi$
\begin{eqnarray}
\mbox{Tr}\partial_{\mu} U \partial^{\mu} U^{\dagger} &=&
\mbox{Tr} {i\over F_\pi} \vec{\tau}\cdot\partial_{\mu}\vec{\phi} \times
{-i\over F_\pi}\vec{\tau}\cdot\partial^{\mu}\vec{\phi}= {2\over F_\pi^2}
\partial_{\mu}\vec{\phi}\cdot \partial^{\mu}\vec{\phi}\,\nonumber\\
{\rm Tr}(U+U^\dagger)&=&{\rm Tr}(2-{1\over F_\pi^2}\vec{\tau}\cdot
\vec{\phi}\vec{\tau}\cdot\vec{\phi})={\rm const.}-{2\over F_\pi^2}\vec{\phi}\cdot\vec{\phi}
\end{eqnarray}
we reproduce the free pion Lagrangian, as required,

At the SU(3) level, including an appropriately 
generalized chiral symmetry breaking term,
there is even predictive power---one has
\begin{equation}
 {F_\pi^2\over 4} \mbox{Tr} \partial_{\mu} U \partial^{\mu} U^{\dagger}
=   {1\over 2} \sum_{j=1}^8 \partial_{\mu}
\phi_j\partial^{\mu}\phi_j +\cdots \nonumber\\
\end{equation}
\begin{eqnarray}
{F_\pi^2 \over 4} \mbox{Tr} 2 B_0 m ( U+ U^{\dagger})
& =&  \mbox{const.}
-{1\over 2} (m_u+ m_d)B_0 \sum_{j=1}^3 \phi^2_j \nonumber\\
&-&{1\over 4} (m_u+m_d+2m_s)B_0\sum_{j=4}^7 \phi^2_j
 -{1\over 6} (m_u+m_d +4m_s)B_0\phi^2_8  +\cdots \, \nonumber\\
&&
\end{eqnarray}
where $B_0$ is a constant and $m$ is the quark mass matrix. We can
then identify the meson masses as
\begin{eqnarray}
 m^2_{\pi} & =&  2\hat{m} B_0
\nonumber \\
 m_K^2 &=& (\hat{m} +m_s) B_0 \nonumber \\
m_{\eta}^2 & =& {2\over 3} (\hat{m} + 2m_s) B_0  \, ,
\end{eqnarray}
where $\hat{m}={1\over 2}(m_u+m_d)$ is the mean light quark mass.
This system of three equations is {\it overdetermined}, and we find by simple
algebra
\begin{equation}
3m_{\eta}^2 +m_{\pi}^2 - 4m_K^2 =0 \, \, .
\end{equation}
which is the Gell-Mann-Okubo mass relation and is well-satisfied
experimentally\cite{gmo}.  Expanding to fourth order in the fields we 
also reproduce the well-known and experimentally successful
Weinberg $\pi\pi$ scattering lengths\cite{weib}
\begin{equation}
a_0^0={7m_\pi^2\over 32\pi F_\pi^2},\quad a_0^2=-{m_\pi^2\over 16\pi
F_\pi^2}, \quad a_1^1={m_\pi^2\over 24\pi F_\pi^2} 
\end{equation}

However, when one attempts to go beyond tree level in order to unitarize 
the results, divergences arise and that is where the field stopped at the 
end of the 1960's.  The solution, as pointed out ten years later 
by Weinberg\cite{wbp} 
and carried out by 
Gasser and Leutwyler\cite{gl}, is to absorb these 
divergences in phenomenological
constants, just as done in QED.  A new wrinkle in this case is that
the theory is nonrenormalizabile in that the forms of the divergences are
{\it different} from the terms that one started with.  That means that 
the form of the counterterms that are used to absorb these divergences 
must also be different, and Gasser and Leutwyler wrote down the most general
counterterm Lagrangian that one can have at one loop, which involves 
{\it four-derivative} interactions
\begin{eqnarray}
{\cal L}_4 &  =&\sum^{10}_{i=1} L_i {\cal O}_i
= L_1\bigg[{\rm tr}(D_{\mu}UD^{\mu}U^{\dagger})
\bigg]^2+L_2{\rm tr} (D_{\mu}UD_{\nu}U^{\dagger})\cdot
{\rm tr} (D^{\mu}UD^{\nu}U^{\dagger}) \nonumber \\
 &+&L_3{\rm tr} (D_{\mu}U D^{\mu}U^{\dagger}
D_{\nu}U D^{\nu}U^{\dagger})
+L_4 {\rm tr}  (D_{\mu}U D^{\mu}U^{\dagger})
{\rm tr} (\chi{U^{\dagger}}+U{\chi}^{\dagger}
) \nonumber \\
&+&L_5{\rm tr} \left(D_{\mu}U D^{\mu}U^{\dagger}
\left(\chi U^{\dagger}+U \chi^{\dagger}\right)
\right)+L_6\bigg[ {\rm tr} \left(\chi U^{\dagger}+
U \chi^{\dagger}\right)\bigg]^2 \nonumber \\
&+&L_7\bigg[ {\rm tr} \left(\chi^{\dagger}U-
U\chi^{\dagger}\right)\bigg]^2 +L_8 {\rm tr}
\left(\chi U^{\dagger}\chi U^{\dagger}
+U \chi^{\dagger}
U\chi^{\dagger}\right)\nonumber \\
&+&iL_9 {\rm tr} \left(F^L_{\mu\nu}D^{\mu}U D^{\nu}
U^{\dagger}+F^R_{\mu\nu}D^{\mu} U^{\dagger}
D^{\nu} U \right) +L_{10} {\rm tr}\left(F^L_{\mu\nu}
U F^{R\mu\nu}U^{\dagger}\right) \nonumber\\
\end{eqnarray}
where the covariant derivative is defined via
\begin{equation}
D_\mu U=\partial_\mu U+\{A_\mu,U\}+[V_\mu,U]
\end{equation}
the constants $L_i, i=1,2,\ldots 10$ are arbitrary (not determined from chiral
symmetry alone) and
$F^L_{\mu\nu}, F^R_{\mu\nu}$ are external field strength tensors defined via
\begin{eqnarray}
F^{L,R}_{\mu\nu}=\partial_\mu F^{L,R}_\nu-\partial_\nu
F^{L,R}_\mu-i[F^{L,R}_\mu ,F^{L,R}_\nu],\qquad F^{L,R}_\mu =V_\mu\pm A_\mu .
\end{eqnarray}
Now just as in the case of QED the bare parameters $L_i$ which appear
in this Lagrangian are not physical quantities.  Instead the experimentally 
relevant (renormalized)
values of these parameters are obtained by appending to these bare values
the divergent one-loop contributions---
\begin{equation} L^r_i = L_i -{\gamma_i\over 32\pi^2}
\left[{-2\over \epsilon} -\ln (4\pi)+\gamma -1\right]\end{equation}
By comparing predictions with experiment, Gasser and Leutwyler were able 
to determine empirical
values for each of these ten parameters.  
Typical results are shown in Table 1, together with the way in which they
were determined.
\begin{table}
\begin{center}
\begin{tabular}{l l c}\hline\hline
Coefficient & Value & Origin \\
\hline
$L_1^r$ & $0.65\pm 0.28$ & $\pi\pi$ scattering \\
$L_2^r$ & $1.89\pm 0.26$ & and\\
$L_3^r$ & $-3.06\pm 0.92$ & $K_{\ell 4}$ decay \\
$L_5^r$ & $2.3\pm 0.2$ & $F_K/F_\pi$\\
$L_9^r$ & $7.1\pm 0.3$ & $\pi$ charge radius \\
$L_{10}^r$ & $-5.6\pm 0.3$ & $\pi\rightarrow e\nu\gamma$\\
\hline\hline
\end{tabular}
\caption{Gasser-Leutwyler counterterms and the means by which
they are determined.}
\end{center}\label{tbl;a}
\end{table}
The important question to ask at this point is why stop at order 
four derivatives?
Clearly if two-loop amplitudes from ${\cal L}_2$ or one-loop
corrections from ${\cal L}_4$ are calculated, divergences will arise which
are of six-derivative character.  Why not include these?  The answer is that
the chiral procedure represents an expansion in energy-momentum.  Corrections
to the lowest order (tree level) predictions from one-loop 
corrections from ${\cal L}_2$
or tree level contributions from ${\cal L}_4$ are ${\cal
O}(E^2/\Lambda_\chi^2)$
where $\Lambda_\chi\sim 4\pi F_\pi\sim 1$ GeV is the chiral scale\cite{sca}.
Thus chiral
perturbation theory is a {\it low energy} procedure.  It is only to the extent
that the energy is small compared to the chiral scale that it makes sense to
truncate the expansion at the one-loop (four-derivative) level.  
Realistically this
means that we deal with processes involving $E<500$ MeV, and 
for such reactions the procedure is found to work very well.

In fact 
Gasser and Leutwyler, besides giving the form of the ${\cal O}(p^4)$ chiral
Lagrangian, have also performed the one loop integration and have written the
result in a simple algebraic form.  Users merely need to look up the result in
their paper and, despite having ten phenomenological constants the theory is
quite predictive.  An example is shown in Table 2, where predictions are
given involving quantities which arise using just two of the 
constants---$L_9,L_{10}$.  The table also reveals an interesting
dilemma---one solid chiral prediction, that for the charged pion 
polarizability, is possibly violated, although this is far from clear
since there are three experimental results here, only one of which is
in disagreement.  This represents a serious challenge to the chiral
predictions (and therefore to QCD!) and should
be the focus of future experimental work.  However, 
there are no Bates implications and, because of space
limitations, we shall have to be
content to stop here.   Interested readers, however, 
can find applications to this and other
systems in a number of review articles\cite{cptr}.

\begin{table}
\begin{center}
\begin{tabular}{cccc}\hline\hline
Reaction&Quantity&Theory&Experiment\\
\hline
$\pi^+\rightarrow e^+\nu_e\gamma$ & $h_V(m_\pi^{-1})$ & 0.027 
& $0.029\pm 0.017$\cite{pdg}\\
$\pi^+\rightarrow e^+\nu_ee^+e^-$ & $r_V/h_V$ & 2.6 & $2.3\pm 0.6$\cite{pdg}\\
$\gamma\pi^+\rightarrow\gamma\pi^+$ & $(\alpha_E+\beta_M)\,(10^{-4}\,{\rm fm}^3)$& 0
&$1.4\pm 3.1$\cite{anti}\\
      &$\alpha_E\,(10^{-4}\,{\rm fm}^3)$&2.8 & $6.8\pm 1.4$\cite{anti1}\\
 & & & $12\pm 20$\cite{russ}\\
 & & & $2.1\pm 1.1$\cite{slac}\\
\hline
\end{tabular}
\caption{Chiral Predictions and data in radiative pion processes.}
\end{center}\label{tbl;b}
\end{table}

\section{$\chi$pt and Bates}

For application at Bates it is important to note that the same ideas can
be applied within the sector of meson-nucleon interactions, although with 
a bit more difficulty.  
Again much work
has been done in this regard\cite{gss}, but there remain important 
challenges\cite{bkm}.  Writing
the lowest order chiral Lagrangian at the SU(2) level is
straightforward---
\begin{equation}
{\cal L}_{\pi N}=\bar{N}(i\not\!\!{D}-m_N+{g_A\over 2}\rlap /{u}\gamma_5)N
\end{equation}
where $g_A$ is the usual nucleon axial coupling in the chiral limit, the
covariant derivative $D_\mu=\partial_\mu+\Gamma_\mu$ is given by
\begin{equation}
\Gamma_\mu={1\over 2}[u^\dagger,\partial_\mu u]-{i\over 2}u^\dagger
(V_\mu+A_\mu)u-{i\over 2}u(V_\mu-A_\mu)u^\dagger ,
\end{equation}
and $u_\mu$ represents the axial structure
\begin{equation}
u_\mu=iu^\dagger\nabla_\mu Uu^\dagger
\end{equation}
Expanding to lowest order we find
\begin{eqnarray}
{\cal L}_{\pi N}&=&\bar{N}(i\rlap /{\partial}-m_N)N+g_A
\bar{N}\gamma^\mu\gamma_5{1\over 2}\vec{\tau}N\cdot({i\over F_\pi}\partial_\mu\vec{\pi}
+2\vec{A}_\mu)\nonumber\\
&-&{1\over 4F_\pi^2}\bar{N}\gamma^\mu\vec{\tau}N\cdot\vec{\pi}\times
\partial_\mu\vec{\pi}+\ldots
\end{eqnarray}
which yields the Goldberger-Treiman relation,
connecting strong and weak couplings of the nucleon system\cite{gt}
\begin{equation}
F_\pi g_{\pi NN}=m_N g_A
\end{equation}
Using the present best values for these quantities, we find
\begin{equation}
92.4 \mbox{MeV}\times 13.05 =1206 \mbox{MeV}\quad\mbox{vs.}\quad 1189 \mbox{MeV}
= 939\mbox{MeV}\times 1.266
\end{equation}
and the agreement to better than two percent strongly confirms the validity
of chiral symmetry in the nucleon sector.  Actually the Goldberger--Treiman relation
is only strictly true at the unphysical point $g_{\pi NN}(q^2=0)$ and 
one {\it expects}
about a ~1\% discrepancy to exist.  An interesting "wrinkle" in this regard
is the use of the so-called Dashen-Weinstein relation, which takes
into account lowest order SU(3)
symmetry breaking, to predict this discrepancy in terms of corresponding numbers in the
strangeness changing sector\cite{dw}.

Another successful application at tree level involves threshold charged 
pion photoproduction and the Kroll-Ruderman term\cite{kr}, which arises 
from the feature that, since the pion must be derivatively coupled, 
there exists a $\bar{N}N\pi^\pm\gamma$ contact interaction which dominates
threshold charged pion photoproduction.  Here what is measured is 
the s-wave or $E_{0+}$ multipole, defined via
\begin{equation}
\mbox{Amp}=4\pi(1+\mu)E_{0+}\vec{\sigma}\cdot\hat{\epsilon}+\ldots
\end{equation}
where $\mu=m_\pi/M$. 
The chiral symmetry prediction is\cite{krth}
\begin{eqnarray}
E_{0+}&=&\pm{1\over 4\pi(1+\mu)}{eg_A\over \sqrt{2}F_\pi}(1\mp{\mu\over 2})=
{eg_A\over 4\sqrt{2}F_\pi}\left(\begin{array}{cc}
1-{3\over 2}\mu & \pi^+ \\
-1+{1\over 2}\mu & \pi^-
\end{array}\right)\nonumber\\
&=&\left\{\begin{array}{ll}
+26.3 \times 10^{-3}/m_\pi  & \pi^+n \\
-31.3 \times 10^{-3}/m_\pi  & \pi^-p
\end{array}\right.,
 \end{eqnarray}
which is in excellent agreement with the
present experimental results, as shown in Table 3.
\begin{table}
\begin{center}
\begin{tabular}{ll}
Quantity & Expt.\\
$E_{0+}(\gamma p\rightarrow\pi^+n)$&$(+27.9\pm 0.5)\times 
10^{-3}/m_\pi$\cite{burg1}\\
 \quad&$(+28.8\pm 0.7)\times 10^{-3}/m_\pi$\cite{adam}\\
 \quad&$(+27.6\pm 0.3)\times 10^{-3}/m_\pi$\cite{sal}\\
$E_{0+}(\gamma n\rightarrow\pi^-p)$&$(-31.4\pm 1.3)
\times 10^{-3}/m_\pi$\cite{burg1}\\
\quad&$(-32.2\pm 1.2)\times 10^{-3}/m_\pi$\cite{gold1}\\
\quad&$(-31.5\pm 0.8)\times 10^{-3}/m_\pi$\cite{tri}
\end{tabular}
\caption{Experimental values for $E_{0+}$ multipoles in charged pion
photoproduction.}
\end{center}\label{tbl;c} 
\end{table}

  However, any realistic approach must also involve loop calculations 
as well as the use of a Foldy-Wouthuysen transformation in order to assure
proper power counting.  This approach goes under the name of heavy baryon
chiral perturbation theory (HB$\chi$pt) and interested readers can find a 
compendium of such results in the review article\cite{meisrv}.  
For our purposes we shall have to be content to examine just two 
applications.  One is neutral pion photoproduction.  In this case the
Kroll-Ruderman term is absent and the chiral expansion of the $E_{0+}$
threshold amplitude begins at order $\mu$ and a heavy baryon HB$\chi$pt
calculation by Bernard, Kaiser, and Meissner found an important loop
contribution which had been omitted in the previous PCAC/based 
approach\cite{deb}.  The correct chiral prediction at ${\cal O}(\mu^2)$
was found to be\cite{vgkm} 
\begin{equation}
E_{0+}={eg_A\over 8\pi M}\mu\{1-[{1\over 2}(3+\kappa_p)+({M\over 4F_\pi})^2]\mu
+{\cal O}(\mu^2)\}
\end{equation}
where the term in $M^2$ signifies the ``new'' chiral loop contribution.
However, comparison with experiment is tricky because of the existence of
isotopic spin breaking in the pion and nucleon masses, so that
there are {\it two} thresholds---one for $\pi^0p$ and the second for
$\pi^+n$---only
7 MeV apart.  When the physical masses of the pions are used recent data 
from both Mainz and from Saskatoon agree with the
chiral prediction.  However, there are concerns about the convergence of the
chiral expansion, which reads $E_{0+}=C(1-1.26+0.59+\ldots)$. 
There also exist chiral predictions for threshold
p-wave amplitudes which are in good agreement with experiment, 
as shown in Table 4, and 
for which the convergence is exprcted to be rapid.

\begin{table}
\begin{center}
\begin{tabular}{ccc}
 & theory&expt. \\
$E_{0+}(\pi^0p)(\times 10^{-3}/m_\pi)$& -1.2&$ -1.31\pm 0.08$\cite{mai}\\
 & & $-1.32\pm 0.11$\cite{salp}\\
$E_{0+}(\pi^0n)(\times 10^{-3}/m_\pi)$& 2.1& $1.9\pm 0.3$\cite{arg}\\
$P_1/|\vec{q}|(\pi p)(\times{\rm GeV}^{-2})$&0.48&$0.47\pm 0.01$\cite{mai}\\
 & &$0.41\pm 0.03$\cite{salp}
\end{tabular}
\caption{Threshold parameters for neutral pion photoproduction.}
\end{center}\label{tbl;d}
\end{table}

Finally exists a chiral symmetry prediction for the reaction $\gamma n
\rightarrow \pi^0n$
\begin{equation}
E_{0+}=-{eg_A\over 8\pi M}\mu^2\{{1\over 2}\kappa_n+({M\over 4F_\pi})^2\}+\ldots=
2.13\times 10^{-3}/m_\pi
\end{equation}
However, the experimental measurement of such an amplitude involves
considerable challenge, and must be accomplished either by use of a deuterium
target with the difficult subtraction of the proton contribution and of
meson exchange contributions or by use of a ${}^3$He target.  
Neither of these 
are straightforward although some limited data already exist\cite{arg}.

Our final example involves an experiment at Bates---measurement of 
the {\it generalized} proton {\it polarizability} 
via virtual Compton scattering.  First recall from section 2 
the concept of polarizability as the constant of
proportionality between an applied electric or magnetizing field and the
resultant induced electric or magnetic dipole moment---
\begin{equation}
\vec{p}=4\pi\alpha_E\vec{E},\qquad \vec{\mu}=4\pi\beta_M\vec{H}
\end{equation}
The corresponding interaction energy is
\begin{equation}
E=-{1\over 2}4\pi\alpha_EE^2-{1\over 2}4\pi\beta_MH^2
\end{equation}
which, upon quantization, leads to a proton Compton scattering cross section
\begin{eqnarray}
{d\sigma\over d\Omega}&=&\left({\alpha_{em}\over m}\right)^2\left({\omega'\over
\omega}\right)^2[{1\over 2}
(1+\cos^2\theta)\nonumber\\
&-&{m\omega\omega'\over \alpha_{em}}[{1\over
2}(\alpha_E+\beta_M)(1+\cos\theta)^2
+{1\over 2}(\alpha_E-\beta_M)(1-\cos\theta)^2+\ldots].\label{eq:sss}
\end{eqnarray}
It is clear from Eq.(\ref{eq:sss})
that, from careful measurement of the differential scattering cross section,
extraction of these structure dependent polarizability terms is possible
provided that \begin{itemize}
\item [i)] the energy is large enough that these terms are significant compared to the
leading
Thomson piece and 
\item [ii)] that the energy is not so large that higher order
corrections become important
\end{itemize} 
and this has been accomplished recently at
SAL and MAMI, yielding\cite{protpol}
\begin{equation}
\alpha_E^{exp}=(12.1\pm 0.8\pm 0.5)\times 10^{-4}\mbox{fm}^3,\qquad
\beta_M^{exp}=(2.1\mp 0.8\mp 0.5)\times
10^{-4}\mbox{fm}^3
\end{equation}
A chiral one loop calculation has also been performed by Bernard, Kaiser, and
Meissner and yields a result in good agreement with these 
measurements\cite{bkmt}
\begin{equation}
\alpha_E^{theo}=10\beta_M^{theo}={5e^2g_A^2\over 384\pi^2F_\pi^2m_\pi}=
12.2\times 10^{-4}\mbox{fm}^3
\end{equation}
The idea of {\it generalized} polarizability can be understood from the
analogous venue of electron scattering wherein measurement of the charge
form factor as a function of $\vec{q}^2$ leads, when Fourier transformed,
to a picture of the {\it local} charge density within the system.  In the
same way the virtual Compton scattering process---$\gamma^*+p\rightarrow
\gamma+p$ can provide a measurement of the $\vec{q}^2$-dependent 
electric and magnetic polarizabilities, whose Fourier transform
provides a picture of the {\it local
polarization density} within the proton.  On the theoretical side our group 
has performed a one loop HB$\chi$pt calculation and has produced a closed from
expression for the predicted polarizabilities\cite{hemm}
 \begin{eqnarray}
\bar{\alpha}^{(3)}_E(\bar{q})&=& \frac{e^2 g_{A}^2 m_\pi}{64\pi^2
F_{\pi}^2}\;\frac{4+2\frac{\bar{q}^2}{m_{\pi}^2}-\left(8-2\frac{\bar{q}^2}{m_{\pi}^2}
-\frac{\bar{q}^4}{m_{\pi}^4}\right)\frac{m_\pi}{\bar{q}}\arctan\frac{\bar{q}}{2
m_{\pi}}}{\bar{q}^2\left(4+\frac{\bar{q}^2}{m_{\pi}^2}\right)}
\; , \nonumber\\
\bar{\beta}^{(3)}_M(\bar{q})&=& \frac{e^2 g_{A}^2 m_\pi}{128\pi^2
F_{\pi}^2}\;\frac{-\left(4+2\frac{\bar{q}^2}{m_{\pi}^2}\right)+\left(8+6\frac{
\bar{q}^2}{m_{\pi}^2}+\frac{\bar{q}^4}{m_{\pi}^4}\right)\frac{m_\pi}{\bar{q}}\arctan
\frac{\bar{q}}{2 m_{\pi}}}{\bar{q}^2\left(4+\frac{\bar{q}^2}{m_{\pi}^2}\right)}
\; . \label{eq:bq}
\end{eqnarray}
In the electric case the structure is about what would be expected---a gradual
falloff of $\alpha_E(\bar{q})$ from the real photon point with scale
$r_p\sim m_\pi$.  However, the magnetic generalized polarizability is
predicted to {\it rise} before this general falloff occurs---chiral
symmetry requires the presence of both a paramagnetic and a diamagnetic
component to the proton.  Both predictions have received some
support in a soon to be announced (and tour de force) MAMI measurement
at $\bar{q}=600$ MeV\cite{mami}.  However, 
since parallel kinematics were employed in the experiment
the desired generalized polarizabilities had to be identified on top
of an enormous Bethe-Heitler background.  The Bates measurement, 
to be performed by
the OOPS collaboration next spring, will take place at $\bar{q}=240$ MeV
and will use the cababilities of the OOPS detector system to provide
a 90 degree out of plane measurement, which should be {\it much} less
sensitive to the Bethe-Heitler blowtorch.  We anxiously await the results.

\section{Conclusion}

In a short paper it is not possible to give any sense of the range of 
phenomena to which the concept of effective field theory as manifested via
chiral perturbation theory has been applied, and interested readers can
find many further applications in \cite{cptr} and \cite{meisrv}.  
Nevertheless, we have tried to convey 
the relatively direct connection of such predictions to the
underlying QCD interaction and the feature that in this way QCD itself can
be tested at Bates.  

\begin{center}
{\bf Acknowlegement}
\end{center}

It is a pleasure to acknowledge the hospitality of MIT/Bates and the organizers
of this meeting.  This work was supported in part by the National Science
Foundation.


\begin{thebibliography}{99}
\bibitem{eftr} See, {\it e.g.} A. Manohar, "Effective Field Theories," in
{\bf Schladming 1966: Perturbative and Nonperturbative 
Aspects of Quantum Field
Theory}, hep-ph/9606222; D. Kaplan, "Effective Field 
Theories," in Proc. 7th Summer
School in Nuclear Physics, nucl-th/9506035,; H. Georgi, "Effective 
Field Theory," in
Ann. Rev. Nucl Sci. {\bf 43}, 209 (1995).
\bibitem{skb} B.R. Holstein, Am. J. Phys. {\bf 67}, 422 (1999).
\bibitem{feyn} A corresponding classical physics discussion is given 
in R.P Feynman,
R.B. Leighton, and M. Sands, {\bf The Feynman Lecures on Physics}, 
Addison-Wesley, Reading, MA, (1963) Vol. I, Ch. 32. 
\bibitem{goldstone} J. Goldstone, Nuovo Cim. {\bf 19}, 154 (1961); J.
Goldstone, A. Salam, and S. Weinberg, Phys. Rev. {\bf 127}, 965 (1962).
\bibitem{gg} S. Gasiorowicz and D.A. Geffen, Rev. Mod. Phys. {\bf 41}, 
531 (1969).
\bibitem{gmo} M. Gell-Mann, CalTech Rept. {\bf CTSL-20} (1961); S. Okubo,
Prog. Theo. Phys. {\bf 27}, 949 (1962).
\bibitem{weib} S. Weinberg, Phys. Rev. Lett. {\bf 17} 616 (1966).
\bibitem{wbp} S. Weinberg, Physica {\bf A96}, 327 (1979).
\bibitem{gl} J. Gasser and H. Leutwyler, Ann. Phys. (NY) {\bf 158}, 142 
(1984); Nucl. Phys. {\bf B250}, 465 (1985).  
\bibitem{sca} A. Manohar and H. Georgi, Nucl. Phys. {\bf B234}, 189 (1984);
J.F. Donoghue, E. Golowich and B.R. Holstein, Phys. Rev. {\bf D30},
587 (1984).
\bibitem{pdg} Particle Data Group, Phys. Rev. {\bf D54}, 1 (1996).
\bibitem{anti} Yu. M. Antipov et al., Z. Phys. {\bf C26}, 495 (1985). 
\bibitem{anti1} Yu. M. Antipov et al., Phys. Lett. {\bf B121}, 445 (1983).
\bibitem{russ} T.A. Aibergenov et al., Czech. J. Phys. {\bf 36}, 948 (1986).
\bibitem{slac} D. Babusci et al., Phys. Lett. {\bf B277}, 158 (1992).
\bibitem{cptr} See, {\it e.g.} B.R. Holstein, Int.J. Mod. Phys. {\bf A7},
7873 (1993); H. Leutwyler, in {\bf Perspectives in the Standard Model}, eds. 
R.K. Ellis, C.T. Hill, and J.D. Lykken, World Scientific, Singapore (1992);
J. Gasser, in Advanced School on Effective Theories, eds. F. Cornet and M.J.
Herrero, World Scientific, Singapore (1997); H. Leutwyler, in {\bf Selected 
Topics in Nonperturbative QCD}, eds. A. DiGiacomo and D. Diakonov, IOS
Press, Amsterdam (1996).
\bibitem{gss} J. Gasser, M. Sainio, and A. Svarc, Nucl. Phys. {\bf B307}, 
779 (1988). 
\bibitem{bkm} V. Bernard, N. Kaiser, and U.G. Meissner, Int. J. Mod. Phys. 
{\bf E4}, 193 (1995).
\bibitem{gt} M. Goldberger and S.B. Treiman, Phys. Rev. {\bf 110}, 1478 (1958).
\bibitem{dw} R. Dashen and M. Weinstein, Phys. Rev. {\bf 188}, 2330 (1969); 
B.R. Holstein, "Nucleon Axial Matrix Elements," 
Few-Body Systems Suppl. {\bf 11}, 116 (1999); J.L. Goity, R. Lewis, 
and M. Schvelinger, "The Goldberger-Treiman Discrepancy in SU(3)," Phys.
Lett. {\bf B454}, 115 (1999).
\bibitem{kr} N. Kroll and M.A. Ruderman, Phys. Rev. {\bf 93}, 233 (1954).
\bibitem{krth} P. deBaenst, Nucl. Phys. {\bf B24}, 613 (1970).
\bibitem{burg1} J.P. Burg, Ann. De Phys. (Paris) {\bf 10}, 363 (1965).
\bibitem{adam} M.J. Adamovitch et al., Sov. J. Nucl. Phys. {\bf 2}, 95 (1966).
\bibitem{sal} J. Bergstrom, private communication. 
\bibitem{gold1} E.L. Goldwasser et al., {\it Proc. XII Int. Conf. on High Energy
Physics, Dubna, 1964}, ed. Ya.-A Smorodinsky, Atomizdat, Moscow (1966).
\bibitem{tri} M. Kovash, $\pi N$ Newsletter {\bf 12}, 51 (1997).
\bibitem{meisrv} V. Bernard, U.-G. Meissner, and N. Kaiser, Int. J. Mod. 
Phys. {\bf E4}, 193 (1995).
\bibitem{deb} P. deBaenst, Nucl. Phys. {\bf B24}, 633 (1970); A.M. Bernstein
and B.R. Holstein, Comm. Nucl. Part. Phys. {\bf 20}, 197 (1991). 
\bibitem{vgkm} V. Bernard, J. Gasser, N. Kaiser and Ulf-G. Meissner, Phys.
Lett. {\bf B268}, 291 (1991).
\bibitem{mai} M. Fuchs et al., Phys. Lett. {\bf B368}, 20 (1996).
\bibitem{salp} J.C. Bergstrom et al., Phys. Rev. {\bf C53}, R1052 (1996). 
\bibitem{arg} P. Argan et al., Phys. Lett. {\bf B206}, 4 (1988). 
\bibitem{protpol} F.J. Federspiel et al., Phys. Rev. Lett. {\bf 67},
1511 (1991); A. L. Hallin et al., Phys. Rev. {\bf C48}, 1497 (1993);
A. Zieger et al., Phys. Lett. {\bf B278}, 34 (1992); B.R. MacGibbon et
al., Phys. Rev. {\bf C52}, 2097 (1995).
\bibitem{bkmt} V. Bernard, N. Kaiser, and U.-G. Meissner, Phys. Rev.
Lett. {\bf 67}, 1515 (1991).
\bibitem{hemm} T.R. Hemmert, B.R. Holstein, G. Knoechlein, and D.
Drechsel, hep-ph/9910036.
\bibitem{mami} S. Kerhoas et al., Few Body Syst. Supp. {\bf 10}, 523 (1999).
\end{thebibliography}
\end{document}